\pdfoutput=1

\documentclass[11pt,twoside,a4paper,cmspaper,final,collab]{cms-tdr}
\def\svnVersion{438865P}\def\cmsCernNoTag{CERN-EP-2017-133}\def\cmsCernDate{\today}\def\cmsMessage{Published in Physical Review C as \href{http://dx.doi.org/10.1103/PhysRevC.96.064902}{\doi{10.1103/PhysRevC.96.064902.}}}

\begin{document}\cmsNoteHeader{HIN-15-010}

\hyphenation{had-ron-i-za-tion}
\hyphenation{cal-or-i-me-ter}
\hyphenation{de-vices}
\RCS$Revision: 302630 $
\RCS$HeadURL: svn+ssh://svn.cern.ch/reps/tdr2/notes/HIN-15-010/trunk/HIN-15-010.tex $
\RCS$Id: HIN-15-010.tex 302630 2015-09-04 15:39:56Z mjovan $
\newlength\cmsFigWidth
\ifthenelse{\boolean{cms@external}}{\setlength\cmsFigWidth{0.49\textwidth}}{\setlength\cmsFigWidth{0.95\textwidth}}
\ifthenelse{\boolean{cms@external}}{\providecommand{\cmsLeft}{top\xspace}}{\providecommand{\cmsLeft}{left\xspace}}
\ifthenelse{\boolean{cms@external}}{\providecommand{\cmsRight}{bottom\xspace}}{\providecommand{\cmsRight}{right\xspace}}
\newcommand {\rootsNN}  {\ensuremath{\sqrt{\smash[b]{s_{_{\mathrm{NN}}}}}}\xspace}
\newcommand {\PbPb}  {\ensuremath{\text{PbPb}}\xspace}
\newcommand {\pPb}  {\ensuremath{\text{pPb}}\xspace}
\newcommand{\Pb}{\ensuremath{\mathrm{Pb}}\xspace}
\renewcommand{\Pp}{\ensuremath{\mathrm{p}}\xspace}
\providecommand{\HYDJET} {\textsc{hydjet}\xspace}
\providecommand{\EPOS} {\textsc{epos}\xspace}

\cmsNoteHeader{HIN-15-010}
\title{Principal-component analysis of two-particle azimuthal correlations in \PbPb and \pPb collisions at CMS}

\date{\today}

\abstract{
For the first time a principle-component analysis is used to separate out
different orthogonal modes of the two-particle correlation matrix from heavy ion collisions.
The analysis uses data from $\rootsNN = 2.76$\TeV \PbPb and $\rootsNN = 5.02$\TeV \pPb collisions collected by the CMS experiment at the LHC.
Two-particle
azimuthal correlations have been extensively used to study hydrodynamic flow in
heavy ion collisions. Recently it has been
shown that the expected factorization of two-particle results into a product of the
constituent single-particle anisotropies is broken. The new information provided by these modes may shed light
on the breakdown of flow factorization in heavy ion collisions.
The first two modes (``leading" and
``subleading") of two-particle correlations are presented for elliptical and triangular
anisotropies in \PbPb and \pPb collisions as a function of \pt over a wide range of event
activity. The leading mode is found to be essentially equivalent to the anisotropy harmonic
previously extracted from two-particle correlation methods. The subleading
mode represents a new experimental observable and is shown to account for a large
fraction of the factorization breaking recently observed at high transverse momentum. The
principle-component analysis technique has also been applied to multiplicity fluctuations.
These  also show a
subleading mode. The connection of these new results to previous studies of factorization
is discussed.
}

\hypersetup{%
pdfauthor={CMS Collaboration},%
pdftitle={Principal-component analysis of two-particle azimuthal correlations in PbPb and pPb collisions at CMS},%
pdfsubject={CMS},%
pdfkeywords={CMS, heavy ion, principal component analysis, initial-state fluctuations, hydrodynamics flow}}

\maketitle
\section{Introduction}

The primary goal of experiments with heavy ion collisions at ultra-relativistic
energies is to study nuclear matter under extreme conditions. Quantum
chromodynamics on the lattice predicts the formation of a quark-gluon plasma (QGP) at energies densities that are attainable in relativistic heavy ion collisions.
Measurements carried out at the Relativistic Heavy Ion Collider (RHIC)
indicate that a strongly interacting QGP is produced in heavy ion collisions \cite{BRAHMS,PHOBOS,STAR,PHENIX}.
The presence of azimuthal anisotropy in the emission of final state hadrons revealed a strong collective flow behavior of this strongly coupled hot and dense medium
\cite{Ollitrault:1992bk,Kolb:2000sd}.
The significantly higher energies available at the CERN LHC compared to RHIC have allowed the ALICE, ATLAS, and CMS experiments to make very detailed measurements of the QGP properties
\cite{Aamodt:2011by,Abelev:2014pua,ATLAS:2012at,Aad:2013xma,Aad:2014fla,Chatrchyan:2012wg,Chatrchyan:2012ta,Chatrchyan:2013kba,HIN_12_011}.
The collective expansion of the QGP can be described by hydrodynamic flow
models \cite{hydro_model1,hydro_model2, hydro_model3}. In the context of these models, the azimuthal anisotropy of hadron emission is the response to the initial density profile of the overlap region of the colliding nuclei.
Such anisotropic emission, for a given event, can be quantified through a Fourier
decomposition of the single-particle distribution
\begin{equation}
\frac{\rd{}N}{d\boldsymbol{p}}=\sum_{n=-\infty}^{\infty}{V_{n}(p)}{\re^{-in\phi}},
\label{flow_eq1}
\end{equation}
with $V_n(p)=v_{n}(p){\re^{in\Psi_n(p)}}$, $\rd\pmb{p}=\rd{\pt}\,{\rd\phi}\,\rd\eta$, and $p$ being a shorthand notation for \pt and $\eta$.
This single-particle distribution is the invariant yield of emitted particles $N$ expressed in
phase space \pt, $\eta$ and $\phi$, \ie, transverse momentum, pseudorapidity, and azimuthal angle.
Here, $v_{n}$ corresponds to the real single-particle anisotropy and
$\Psi_n(p)$ represents the $n$th order event plane angle.
Also, because of the reflection symmetry of the overlap region, the relation $V^{*}_{n} = V_{-n}$ holds for the complex harmonics. Using this relation and
integrating Eq.~(\ref{flow_eq1}) over a given pseudorapidity and \pt window yields
\begin{equation}
\frac{\rd{}N}{\rd{\phi}}=\frac{N}{2\pi}\Big(1+2\sum_{n=1}^{\infty}
v_{n}(p)\cos[n(\phi-\Psi_{n}(p))]\Big).
\label{flow_eq2}
\end{equation}
Note that the single-particle anisotropy coefficient $v_n$ is generally a function of \pt and $\eta$, which is also the case for the event plane angle.
The azimuthal correlation of $N^{\text{pairs}}$ emitted particle pairs (with particles labeled $a$ and $b$) as a
function of their azimuthal separation $\Delta\phi^{ab}=\phi^a-\phi^b$ can be characterized by its own Fourier harmonics
\begin{equation}
\frac{\rd{}N^{\text{pairs}}}{\rd{\Delta\phi^{ab}}}=\frac{N^{\text{pairs}}}{2\pi}\Big(1+2\sum_{n=1}^{\infty}V_{n\Delta}(p^{a},p^{b})\cos(n\Delta\phi)\Big),
\label{flow_eq3}
\end{equation}
where $V_{n\Delta}$ is the two-particle harmonic. In a pure hydrodynamic picture, as a consequence of independent particle emission, the flow hypothesis connects the single- and two-particle spatial anisotropies from Eqs.~(\ref{flow_eq2}) and (\ref{flow_eq3})
through factorization. In other words, particles carry information only about their orientation with respect to the whole system and the two-particle distribution can therefore be factorized based on
\begin{equation}
\langle{\frac{\rd{}N^{\text{pairs}}}{\rd\Delta\phi^{ab}}}\rangle= \langle\frac{\rd{}N}{\rd{\phi^a}}\frac{\rd{}N}{\rd{\phi^b}}\rangle,
\label{eq4}
\end{equation}
with the bracket $\langle{}\rangle$ representing the average over all
events of interest. This equality can be investigated by looking at the connection between the single- and two-particle harmonics
\ifthenelse{\boolean{cms@external}}{
\begin{multline}
{\langle}V_{n\Delta}(p^{a},p^{b}){\rangle}=\langle{{V_n(p^a)}{V^{*}_n(p^b)}}\rangle\\
=\langle{{\upsilon^a_n}{\upsilon^{b}_n}\cos{[n(\Psi^a_n-\Psi^b_n)]}}\rangle\leq\langle{{\upsilon^a_n}{\upsilon^{b}_n}}\rangle.
\label{eq5}
\end{multline}
}{
\begin{equation}
{\langle}V_{n\Delta}(p^{a},p^{b}){\rangle}=\langle{{V_n(p^a)}{V^{*}_n(p^b)}}\rangle=\langle{{\upsilon^a_n}{\upsilon^{b}_n}\cos{[n(\Psi^a_n-\Psi^b_n)]}}\rangle\leq\langle{{\upsilon^a_n}{\upsilon^{b}_n}}\rangle.
\label{eq5}
\end{equation}
}
From Eq.~(\ref{eq5}) we infer that factorization is preserved when the cosine value equals unity. This scenario is possible only when the event plane angle acts as a global phase, lacking any \pt or $\eta$ dependence for a given event. Thus, measurements of the momentum space fluctuations (correlations) constrain the initial state and properties of QGP expansion dynamics. Previous measurements have shown a significant breakdown of factorization at high \pt in ultracentral (\ie, almost head-on) \PbPb collisions~\cite{HIN_12_011}. A smaller effect was also seen in high-multiplicity \pPb collisions~\cite{Khachatryan:2015oea}. Furthermore, significant factorization breakdown effects as a function of $\eta$ were observed in both \PbPb and high-multiplicity \pPb collisions~\cite{Khachatryan:2015oea}. Several possible explanations for the observed factorization breaking have been proposed. One expected contribution arises from nonflow effects, \ie short-range correlations mainly due to jet fragmentation and resonance decays. However, factorization breaking is also possible in hydrodynamic models, once the effects of event-by-event initial-state fluctuations are taken into account~\cite{Gardim:2012im,Heinz:2013bua}. Such a nonuniform initial-state energy density can arise from fluctuations in the positions of nucleons within nuclei and/or the positions of quark and gluon constituents inside each nucleon, giving rise to variations in the collision points when the two nuclei collide. The resulting fluctuating initial energy density profile creates nonuniformities in pressure gradients which push particles in different regions of phase space in directions that vary randomly about a mean angle, thereby imprinting these fluctuations on the final particle distributions. Consequently, the event plane angles estimated from particles in different \pt and $\eta$ ranges may vary with respect to each other. By introducing such a dependence, $\Psi_n=\Psi_n(\pt,\eta)$, it is possible to describe the resulting final-state particle distributions using hydrodynamical models ~\cite{Gardim:2012im,Heinz:2013bua}.

Principal-component analysis (PCA) is a multivariate technique that can separate out the different orthogonal contributions (also known as modes) to the fluctuations.
Using the method introduced in Ref.~\cite{PCA_PUB1}, this paper presents the first experimental use of applying PCA to two-particle correlations in order to study factorization breaking as a function of \pt. This allows the extraction of a new experimental observable, the subleading mode, which is directly connected to initial-state fluctuations and their effect on factorization breaking.

\section{Experimental setup and data samples}
\label{sec:exp_evt}

The Compact Muon Solenoid (CMS) is an axially symmetric detector with an
onion-like structure, which consists of several subsystems concentrically placed around the interaction point.
The CMS magnet is a superconducting solenoid
providing a magnetic field of 3.8\unit{T}, which
allows precise measurement of charged particle momentum. The
muon chambers are placed outside the solenoid. In this analysis the data used is extracted from the silicon tracker, which is the closest subdetector
to the interaction point.
This detector consists of 1440 silicon pixel and
15\,148 silicon strip detector modules that detect hit locations, from which the charged particle trajectories
are reconstructed. The silicon tracker
covers charged particles within the range $\abs{\eta}<
2.5$, and provides an impact parameter resolution of ${\sim}15\mum$ and a
\pt resolution better than 1.5\% up to $\pt{\sim}100\GeVc$.

The other two subdetectors located inside the solenoid, are the
electromagnetic calorimeter (ECAL) and hadronic calorimeter (HCAL). The ECAL
is constructed  of 75\,848 lead-tungstate crystals which are arranged in a
quasi-projective geometry and cover a pseudorapidity range of
$\abs{\eta} < 1.48$ units in the barrel and two endcaps that extend $\abs{\eta}$
up to 3.0. The HCAL barrel and endcaps are sampling calorimeters
constructed from brass and scintillator plates, covering $\abs{\eta} < 3.0$.
Additional extension in $\abs{\eta}$ from 2.9 up to 5.2 is achieved with
the iron and quartz-fiber \v Cerenkov Hadron Forward (HF) calorimeters on
either side of the interaction region.
The HF calorimeters are segmented into towers, each of which is a two-dimensional
cell with a granularity of 0.175$\times$0.175$\unit{rad}^2$ ($\Delta\eta{\times}\Delta\phi$). The zero-degree calorimeters (ZDC) are tungsten-quartz Cherenkov
calorimeters located at $\pm$140\mm from the interaction point~\cite{Grachov:2008qg}.
They are designed to measure the energy of photons and spectator neutrons
emitted from heavy ion collisions.
A set of scintillator tiles, the beam scintillator counters (BSC), are mounted
on the inner side of the HF calorimeters and are used for triggering and beam-halo rejection.
The BSCs cover the range $3.23 < \abs{\eta} < 4.65$.
A detailed description of the CMS detector can be found in
Ref.~\cite{Chatrchyan:2008aa}.

This analysis is performed using data recorded by the CMS experiment during the LHC heavy ion runs in 2011 and 2013.
The \PbPb data set at a center-of-mass energy of $\rootsNN$ = 2.76\TeV corresponds to an
integrated luminosity of about 159\mubinv, while the \pPb data set at $\rootsNN$
= 5.02\TeV corresponds to about 35\nbinv.
During the \pPb run, the beam energies were 4\TeV for protons and 1.58\TeV per nucleon for lead nuclei.

\section{Selection of events and tracks}

Online triggers, track reconstruction, and offline event selections are the same
as in Refs. \cite{HIN_12_011, Chatrchyan:2013nka, Khachatryan:2015oea} for \PbPb and \pPb data samples, and are summarized in the following sections.

\subsection{The \PbPb data}

Minimum bias \PbPb events were collected using coincident trigger signals from both ends of
the detector in either BSCs or the HF calorimeters. Events affected by cosmic rays, detector noise,
out-of-time triggers, and beam backgrounds were suppressed by requiring a coincidence of the
minimum bias trigger with bunches colliding in the interaction region. The efficiency of the trigger is more than 97\%
in case of hadronic inelastic \PbPb collisions. Because of hardware limits on the
data acquisition rate, only a small fraction (2\%) of all minimum bias events were recorded (\ie,
the trigger is ``prescaled"). To enhance the event sample for very central \PbPb collisions, a dedicated
online trigger was implemented by simultaneously requiring the HF transverse energy
(\ET) sum to be greater than 3260\GeV and the pixel cluster multiplicity to be greater than 51400
(which approximately corresponds to 9500 charged particles over 5 units of $\eta$).
The selected events correspond to the 0--0.2\% most central \PbPb collisions. Other standard \PbPb
centrality classes presented in this paper were determined based on the total energy deposited
in the HF calorimeters \cite{Chatrchyan:2012ta}. The inefficiencies of the minimum bias trigger and event selection
for very peripheral events are taken into account.

In order to reduce further the background from single-beam interactions (\eg, beam gas and beam halo),
cosmic muons, and ultraperipheral collisions leading to the electromagnetic breakup
of one or both \Pb nuclei \cite{PhotonuclearExcitations}, offline \PbPb event selection criteria \cite{Chatrchyan:2012ta} were applied by requiring
energy deposits in at least three towers in each of the HF calorimeters, with at least 3\GeV of
energy in each tower, and the presence of a reconstructed primary vertex built of at least
two tracks. The reconstructed primary vertex is required to be located within $\pm$15\unit{cm} of the
average interaction point along the beam axis and within a radius of 0.2 cm in the transverse
plane. Following the procedure developed in Ref.~\cite{HIN_12_011}, events with large signals in both ZDCs and HFs are identified as having at least one additional interaction,
or pileup events, and are thus rejected (about 0.1\% of all events).

The reconstruction of the primary event vertex and of the trajectories of charged particles in
PbPb collisions is based on signals in the silicon pixel and strip detectors and is described in
detail in Ref. \cite{Chatrchyan:2012ta}. From studies based on \PbPb events simulated using \HYDJET v1.8 \cite{JetModel}, the
combined geometrical acceptance and reconstruction efficiency of the primary tracks is about
70\% at $\pt\sim1$\GeVc and $\abs{\eta}<1.0$ for the most central (0--5\%) \PbPb events, but drops to about
50\% for $\pt\sim0.3$\GeVc. The fraction of misidentified tracks is kept to be $<$5\% over
most of the \pt ($>$0.5\GeVc) and $\abs{\eta}$ ($<$1.6) ranges. It increases to about 20\% for very low \pt
($<$0.5\GeVc) particles in the forward ($\abs{\eta}\geq 2.0$) region.

\subsection{The \pPb data}

Minimum bias \pPb events
were triggered by requiring at least one track with $\pt > 0.4$\GeVc to be
found in the pixel tracker in coincidence with an LHC \pPb bunch crossing. From all minimum bias triggered events,
only a fraction of (${\sim}10^{-3}$) was recorded.
In order to select high-multiplicity
\pPb collisions, a dedicated trigger was implemented using
the CMS level-1 (L1) and high-level trigger (HLT) systems. At L1, the total transverse
energy summed over the ECAL and HCAL is required to be greater than a given
threshold (20 or 40\GeV). The online track reconstruction for the HLT is based on the three layers of
pixel detectors, and requires a track originated within a cylindrical region of
length 30\unit{cm} along the beam and radius of 0.2\unit{cm} perpendicular to the beam. For each event,
the vertex reconstructed with the highest number of pixel
tracks is selected. The number of pixel tracks (${N}_\text{trk}^\text{online}$)
with $\abs{\eta}<2.4$, $\pt > 0.4$\GeVc, and having a distance of closest approach of
0.4\unit{cm} or less to this vertex, is determined for each event.

In the offline analysis, hadronic \pPb collisions are selected by
requiring a coincidence of at least one HF calorimeter tower with more than 3\GeV of
total energy in each of the HF detectors.
Events are also required to contain at least one reconstructed primary
vertex within 15\unit{cm} of the nominal interaction point along the beam axis
and within 0.15\unit{cm} transverse to the beam trajectory.
At least two reconstructed tracks are required to be associated
with the primary vertex. Beam-related background is suppressed
by rejecting events for which less than 25\% of all reconstructed tracks
are of good quality (\ie, the tracks selected for physics analysis).

The instantaneous luminosity provided by the LHC
in the 2013 \pPb run resulted in approximately 3\% probability of at least one
additional interaction occurring in the same bunch crossing, \ie
pileup events. Pileup was rejected using a procedure based on the number
 of tracks in a given vertex and the distance between that an additional vertex (see
Ref.~\cite{Chatrchyan:2013nka}).
The fraction of \pPb events selected by these criteria, which have at least one particle (proper lifetime $\tau>10^{-18}\unit{s}$) with total energy $E>3$\GeV in
$\eta$ range of $-5<\eta<-3$ and at least one in the range $3<\eta<5$ (selection referred to as ``double-sided")
has been found to be 97--98\% by using the \EPOS~\cite{EPOS} and \HIJING~\cite{HIJING} event generators.

In this analysis, the CMS {highPurity}~\cite{CMS-PAS-TRK-10-001} tracks are used. Additionally,
a reconstructed track is only considered as a primary-track candidate if
the significance of the separation along the beam axis ($z$) between
the track and the best vertex, $dz/\sigma(dz)$, and the
significance of the impact parameter relative to the best vertex transverse
to the beam, $d_{\mathrm{T}}/\sigma(d_{\mathrm{T}})$, are less than 3 in each case. The
relative uncertainty of the \pt measurement, $\sigma(\pt)/\pt$, is
required to be less than 10\%. To ensure high tracking efficiency and to
reduce the rate of misidentified tracks, only tracks within $\abs{\eta}<2.4$
and with $\pt > 0.3$\GeVc are used in the analysis. The entire \pPb data set is
divided into classes of reconstructed track
multiplicity, $N^\text{offline}_{\text{trk}}$, where primary tracks with $\abs{\eta}<2.4$
and $\pt >0.4$\GeVc are counted. The multiplicity classification in this
analysis is identical to that used in Ref.~\cite{Chatrchyan:2013nka}, where more
details are provided.

\section{Analysis technique}
\label{sec:analysis}

This analysis uses two-particle correlations and PCA as a new flow method that can make use of all
	the information contained in $V_{n\Delta}$ harmonics.  Averaging Eq.~(\ref{flow_eq3}) over all events of interest, within a given reference bin $\pt^{\text{ref}}$, and assuming
	factorization, one can write
	\begin{linenomath}
	\begin{equation}
	\langle{\frac{\rd{}N^{\text{pairs}}}{\rd\Delta\phi}}\rangle=\frac{{\langle}N^{\text{pairs}}\rangle}{2\pi}\Big(1+\sum_{n=1}^{\infty}\upsilon^2_n\{2\}\cos({n}\Delta\phi)\Big),
	\label{ref_flow}
	\end{equation}
	\end{linenomath}
	where $\upsilon_n\{2\}$ is the integrated reference flow calculated from the $V_{n\Delta}$ as
	\begin{linenomath}
	\begin{equation}
	\upsilon_n\{2\}=\frac{\sqrt{V_{n\Delta}(\pt^{\text{ref}},\pt^{\text{ref}})}}{\sqrt{V_{0\Delta}(\pt^{\text{ref}},\pt^{\text{ref}})}},
	\end{equation}
	\end{linenomath}
	with,
	\begin{linenomath}
	\begin{equation}
	V_{n\Delta}(\pt^{\text{ref}}, \pt^{\text{ref}})\equiv\langle\sum_{i\in{\text{ref}}}{\cos({n}\Delta\phi_i})\rangle.
	\label{perfect_Vn}
	\end{equation}
	\end{linenomath}
	Here, the label $V_{0\Delta}$ for $N^{\text{pairs}}$ is used, since the sum over cosine counts the number of pairs for the $n = 0$ case. Calculating the differential flow one gets
	\begin{linenomath}
	\begin{equation}
	\upsilon_n(\pt){\{2\}}\upsilon_n\{2\}=\frac{V_{n\Delta}(\pt,\pt^{\text{ref}})}{V_{0\Delta}(\pt,\pt^{\text{ref}})},
	\end{equation}
	\end{linenomath}
	or,
	\begin{linenomath}
	\begin{equation}	\upsilon_n(\pt)=\frac{V_{n\Delta}(\pt,\pt^{\text{ref}})}{\sqrt{V_{n\Delta}(\pt^{\text{ref}},\pt^{\text{ref}})}}\frac{\sqrt{V_{0\Delta}(\pt^{\text{ref}},\pt^{\text{ref}})}}{V_{0\Delta}(\pt,\pt^{\text{ref}})}.
	\label{single_vn}
	\end{equation}
	\end{linenomath}

	The single-particle anisotropy definition in Eq.~(\ref{single_vn}) includes the $V_{0\Delta}$ terms
	to compensate for the fact that the $V_{n\Delta}$ Fourier harmonics are calculated
        without per-event normalization
        by the number of pairs in the given bin~\cite{HIN_12_011,Khachatryan:2015oea}. This way of
	calculating the cosine term is essential for the PCA to work, since it gives a weight to a bin that is of the order of the number of
	particles in it \cite{PCA_PUB1}.

	In a realistic experiment, the $V_{n\Delta}$ harmonics of Eq.~(\ref{perfect_Vn}) are affected by imperfections in the detector and take the following operational definition
\ifthenelse{\boolean{cms@external}}{
	\begin{linenomath}
	\begin{multline}
	V_{n\Delta}(\pt^a, \pt^b)=\langle{\cos({n\Delta\phi})}\rangle_S-\langle{\cos({n\Delta\phi})}\rangle_B,\\
  n=1,2,3,\ldots.
	\label{V_nD}
	\end{multline}
	\end{linenomath}
}{
	\begin{linenomath}
	\begin{equation}
	V_{n\Delta}(\pt^a, \pt^b)=\langle{\cos({n\Delta\phi})}\rangle_S-\langle{\cos({n\Delta\phi})}\rangle_B,\quad  n=1,2,3,\ldots.
	\label{V_nD}
	\end{equation}
	\end{linenomath}
}
	Here, the first term in the right-hand side of Eq.~(\ref{V_nD}), $\langle{\cos{(n\Delta\phi)}}\rangle_S$, is the two-particle anisotropic signal where the correlated particles belong to the same event. The second term, $\langle{\cos{(n\Delta\phi)}}\rangle_B$
        is a background term that accounts for the nonuniform acceptance of the detector. This term is usually two orders of magnitude smaller than the corresponding signal. It is estimated by mixing particle tracks from two random events.
        These two events have the same 2 cm wide range of the primary vertex position in the $z$ direction and belong to the same
	centrality (track multiplicity) class. For both terms, in order to suppress nonflow correlations, a pseudorapidity
        difference requirement between the two tracks $\abs{\Delta\eta}>2$ is applied.

	\subsection{Factorization breaking}

	The PCA is a multivariate analysis that orders the fluctuations in the data by size. The ordering is done through principal components that represent
	orthogonal eigenvectors of the corresponding covariance data matrix. In the context of flow fluctuations, the components should reveal any significant
	substructure caused by the fluctuating initial state geometry of colliding nuclei. Introducing PCA in terms of factorization breaking
	one can write the Pearson correlation coefficient used for measurement of the effect as in Ref. \cite{Khachatryan:2015oea}
\ifthenelse{\boolean{cms@external}}{
	\begin{linenomath}
	\begin{multline}
	r_n(\pt^a,\pt^b)\equiv\frac{V_{n\Delta}(\pt^a,\pt^b)}{\sqrt{V_{n\Delta}(\pt^a,\pt^a)
	V_{n\Delta}(\pt^b,\pt^b)}}\\
\approx{\langle{\cos{n(\Psi(\pt^a)-\Psi(\pt^b))}}\rangle}.
	\label{rn_def}
	\end{multline}
	\end{linenomath}
}{
	\begin{linenomath}
	\begin{equation}
	r_n(\pt^a,\pt^b)\equiv\frac{V_{n\Delta}(\pt^a,\pt^b)}{\sqrt{V_{n\Delta}(\pt^a,\pt^a)
	V_{n\Delta}(\pt^b,\pt^b)}}\approx{\langle{\cos{n(\Psi(\pt^a)-\Psi(\pt^b))}}\rangle}.
	\label{rn_def}
	\end{equation}
	\end{linenomath}
}
	The ratio $r_n$ is approximated by the cosine term, giving unity if the event plane angle is a global phase, as discussed previously. Expressing the ratio
	through the two-particle harmonic in complex form from Eq. (\ref{eq5}), $r_n$ can only be unity if the complex flow coefficient $V_n(\pt)$
	is generated from one initial geometry. For instance, where the initial geometry of the overlap region is defined by some complex eccentricity ($\varepsilon_n$)
        and a fixed real function $f(\pt)$, \ie, $V_n(\pt)=f(\pt)\varepsilon_n$.
	However, if events are described by multiple eccentricities then $r_n$ may be less than unity and the flow pattern displays factorization breaking \cite{PCA_PUB3}.
	This last statement can be generalized by expanding the complex flow coefficient using the principal components ($V^{(1)}_n(\pt)$, $V^{(2)}_n(\pt),\ldots$) as a basis built from a covariance data matrix of given size $N_\alpha\times{N_\alpha}$
\ifthenelse{\boolean{cms@external}}{
	\begin{linenomath}
	\begin{multline}
	V_n(\pt)=\xi^{(1)}_nV^{(1)}_n(\pt)+\xi^{(2)}_nV^{(2)}_n(\pt)\\
+\cdots+\xi^{(N_\alpha)}_nV^{(N_\alpha)}_n(\pt),
	\label{pca_eq1}
	\end{multline}
	\end{linenomath}
}{
	\begin{linenomath}
	\begin{equation}
	V_n(\pt)=\xi^{(1)}_nV^{(1)}_n(\pt)+\xi^{(2)}_nV^{(2)}_n(\pt)+\cdots+\xi^{(N_\alpha)}_nV^{(N_\alpha)}_n(\pt),
	\label{pca_eq1}
	\end{equation}
	\end{linenomath}
}
	where $\xi^{(i)}_n$ are complex uncorrelated variables with zero mean \ie $\langle{\xi^{(i)}_n\xi^{(j)}_n}\rangle=\delta_{ij}$, $\langle{\xi^{(i)}_n}\rangle=0$, and $N_\alpha$ represents the number of \pt differential bins. Therefore, the two-particle harmonics are the building elements of the covariance data matrix $[\hat{V}_{n\Delta}(\pt^a, \pt^b)]_{N_\alpha\times{N_\alpha}}$.

	A covariance matrix is symmetrical and positive semidefinite (\ie, with eigenvalues $\lambda\geq$0). For the flow matrix, the last trait is valid if there are no nonflow contributions and no strong statistical fluctuations \cite{PCA_PUB1}. Now, calculating the two-particle harmonic using  the expansion from Eq. (\ref{pca_eq1}) one gets
	\begin{linenomath}
	\begin{equation}
	V_{n\Delta}(\pt^a, \pt^a)=\sum^{N_\alpha}_{\alpha=1}V^{(\alpha)}_n(\pt^a)V^{(\alpha)}_n(\pt^b).
	\label{sum_alpha}
	\end{equation}
	\end{linenomath}
	Here, the principal components will be referred to as modes \cite{PCA_PUB1, PCA_PUB2, PCA_PUB3}.
        In order to calculate the modes the spectral decomposition is rewritten as:
	\begin{linenomath}
	\begin{equation}
	V_{n\Delta}(\pt^a, \pt^b)=\sum_{\alpha}\lambda^{(\alpha)}\re^{(\alpha)}{(\pt^a)}e^{(\alpha)}{(\pt^b)},
	\label{spec_def}
	\end{equation}
	\end{linenomath}
	which gives:
	\begin{linenomath}
	\begin{equation}
	V^{(\alpha)}_n(\pt)=\sqrt{\lambda^{(\alpha)}}\re^{(\alpha)}(\pt),
	\label{mode_def}
	\end{equation}
	\end{linenomath}
	where $e^{(\alpha)}(\pt)$ are $(\alpha)$ index values of normalized eigenvectors and $\lambda^{(\alpha)}$ eigenvalues that are sorted in a strict decreasing order $\lambda^{(1)}>\lambda^{(2)}>\cdots>\lambda^{(n)}$.
	Eq.~(\ref{sum_alpha}) shows directly that factorization holds only in the case where just one mode is present. If multiple modes are present in the data, Eqs. (\ref{spec_def}) and (\ref{mode_def}) allow to define a normalized orthogonal basis for the total $v_n$ given in Eq.~(\ref{single_vn}). These basis vectors are defined by:
	\begin{linenomath}
	\begin{equation}
	{v^{(\alpha)}_{n}(\pt)\equiv\frac{V^{(\alpha)}_{n}(\pt)}{V^{(1)}_0(\pt)}}.
	\label{subflow1}
	\end{equation}
	\end{linenomath}
	The normalization factor $V^{(1)}_0$ is the first mode that would follow from Eq.~(\ref{mode_def}) using the matrix of the number of pairs, \ie, the matrix of $V_{0\Delta}$ terms. In practice, the mode $V^{(1)}_0$ has a simple physical meaning: it is the average differential multiplicity $\langle{M(\pt)}\rangle$. However, given the pseudorapidity requirement in the correlations, $V^{(1)}_0$ is proportional to $\langle\sqrt{\smash[b]{N^{\text{pairs}}_{\abs{\Delta\eta}>2}(\pt, \pt)}}\rangle$. In order to restore normalization by the average bin multiplicity $\langle{M(\pt)}\rangle$ an intermediate step is made by multiplying the $V_{n\Delta}(\pt^a,\pt^b)$ with:
	\begin{linenomath}
	\begin{equation}
	\zeta=\Big\langle{\frac{V^{\abs{\eta}<2.4}_0(\pt^a, \pt^b)}{V^{\abs{\Delta\eta}>2}_0(\pt^a, \pt^b)}}\Big\rangle,
	\end{equation}
	\end{linenomath}
        $\zeta$ being the mean value of the ratio of number of all pairs and number of pairs after applying
        $\abs{\Delta\eta} > 2$ selection for the given bins. If the $\eta$ distribution of particles did not depend upon \pt then
        $\zeta$($\pt^a$, $\pt^b$) would be constant for all values of $\pt^a$ and $\pt^b$. In fact, $\zeta$ does have a slight dependence
        on $\pt^a$ and $\pt^b$  with a maximum at low values of $\pt^a$ and $\pt^b$. As events become more central the center of gravity of zeta moves to higher \pt values.         Finally, after applying this correction the eigenvalue problem is solved with new matrix elements
	\begin{linenomath}
	\begin{equation}
	{\tilde{V}}_{n\Delta}(\pt^a, \pt^b){\equiv}\zeta{V_{n\Delta}(\pt^a, \pt^b)}.
	\label{xi}
	\end{equation}
	\end{linenomath}
	Eq.~(\ref{subflow1}) then becomes:
	\begin{linenomath}
	\begin{equation}
	v^{(\alpha)}_{n}(\pt)=\frac{{\tilde{V}}^{(\alpha)}_{n}(\pt)}{\langle{{M(\pt)}\rangle}}.
	\label{sub_flow_definition}
	\end{equation}
	\end{linenomath}
	The leading ($\alpha$ = 1) and the subleading ($\alpha$ = 2) normalized modes (for simplicity, term modes will be used)
        can be thought of as new experimental observables.
        Given that the eigenvalues $\lambda^{(\alpha)}$ are strongly ordered, two components typically describe the variance in the harmonic flow to high accuracy.
        The leading mode is strongly correlated with the event plane, and thus is essentially equivalent to the standard definition of the single-particle anisotropic flow, while the subleading mode is uncorrelated
        with the event plane, and thus quantifies the magnitude of the factorization breaking caused by the initial-state fluctuations.

	\subsection{Multiplicity fluctuations}
	\label{sec:multFluct}

	The PCA can also be applied for investigating multiplicity fluctuations in heavy ion collisions. The multiplicity matrix that is used for extraction of the corresponding modes is built from the following matrix elements:
	\begin{linenomath}
	\begin{equation}
	[\hat{M}(\pt^a, \pt^b)]_{N_\alpha\times{N_\alpha}}=\langle{V_{0\Delta}(\pt^a, \pt^b)}\rangle-\langle{M(\pt^a)}\rangle\langle{M(\pt^b)}\rangle,
	\label{V0}
	\end{equation}
	\end{linenomath}
	where the term $V_{0\Delta}(\pt^a, \pt^b)$ represents the number of pairs for the given bins and $M(\pt)$ the given bin multiplicity. Unlike in the flow cases
        $n = 2$, 3, here no
        pseudorapidity requirement $\abs{\Delta\eta}>2$ is applied when correlating tracks. Using the multiplicity matrix the modes defined by Eq.~(\ref{mode_def}) are derived
        and the leading and subleading modes are calculated with Eq.~(\ref{sub_flow_definition}), excluding the multiplication step in Eq.~(\ref{xi}).
The leading mode represents the ``total multiplicity
fluctuations", \ie, if the higher modes are zero, then $v^{(1)}_0$
would approximately be equal to the standard deviation of multiplicity for
the given \pt bin. The reconstructed subleading mode represents a new observable of the multiplicity spectrum.
The multiplicity results in Section \ref{sec:results} represent exploratory studies and are for simplicity only presented for \PbPb.

\section{Systematic uncertainties}
\label{sec:effCorr}

Several sources of possible systematic uncertainties, such as the event selection, the dimension of the matrix, and the effect of the tracking efficiency were investigated.
Among these sources, only the effect of the tracking efficiency had a noticeable influence on the results.
For all the considered cases $n = 0$, 2, 3 the systematic uncertainties were estimated from the full difference between the final result with and without the correction for the tracking efficiency. Each reconstructed track was weighted by the inverse of the efficiency factor, $\varepsilon_{\text{trk}}(\pt, \eta)$, which is a function of transverse momentum and
pseudorapidity. The efficiency weighting factor accounts for the detector acceptance $A(\pt, \eta)$
and the reconstruction efficiency, $E(\pt, \eta)$ ($\varepsilon_{\text{trk}}=A{\,}E$).

From Eqs.~(\ref{mode_def}) and (\ref{sub_flow_definition}) it can be seen that modes are functions of the eigenvectors and eigenvalues, \ie $e$ and $\lambda$, of the matrix, and of the differential multiplicity $M(\pt)$. When the efficiency correction is applied to each track, a completely new matrix is produced and the multiplicity of tracks also increases. The principal components of this new matrix were then calculated and new modes derived. This procedure gives a robust test of how susceptible the modes are to strong changes in ($\lambda$, $e$, $M$). Table 1 summarizes the uncertainties of the subleading mode in the highest bin $2.5<\pt<3.0$\GeVc for both the \pPb and \PbPb cases.
The systematic uncertainties are estimated values and are rounded to the nearest integer.
For the leading mode, systematic uncertainties are significant only for $n = 0$, while for the subleading mode systematic uncertainties are larger for all the cases $n = 0$, 2, 3. In the lower \pt range, for the multiplicity case $n$=0, the systematic uncertainties of the subleading mode are strongly correlated.

\begin{table}[ht]
\centering
\topcaption{Summary of estimated systematic uncertainties relative to the given mode for
the last \pt bin $2.5 < \pt <3.0$\GeVc for \PbPb and \pPb data.}
\newcolumntype{x}{D{,}{\text{--}}{2.2}}
\begin{scotch}{ x cc{c}@{\hspace*{5pt}}cc{c}@{\hspace*{5pt}}cc}
\multicolumn{1}{l}{\PbPb} & \multicolumn{2}{ c }{$n = 2$} &&  \multicolumn{2}{ c }{$n = 3$}  &&  \multicolumn{2}{ c}{$n = 0$}\\
\cline{2-3}\cline{5-6}\cline{8-9}
\multicolumn{1}{l}{Centrality (\%)} & $\alpha=1$  & $\alpha=2$ &&  $\alpha=1$  & $\alpha=2$ &&  $\alpha=1$  & $\alpha=2$\\
0,0.2  & 1\% & 30\%   && 1\%  & 40\%    && 40\% & 10\%  \\
0,5    & 1\%  & 50\%   && 1\%  & 40\%    && 15\% & 10\%  \\
0,10   & 1\% & 30\%   && 1\% & 40\%     && 10\% & 30\%  \\
10,20  & 1\% & 10\%  && 1\% & 40\%     && 10\% & 20\%  \\
20,30  & 1\% & 10\%  && 1\% & 20\%    && 10\% & 15\%  \\
30,40  & 1\% & 10\%  && 1\% & 35\%    && 10\% & 10\%  \\
40,50  & 1\% & 10\%  && 1\% & 25\%    && 10\% & 10\%  \\
50,60  & 1\% & 7\%    && 1\% & 30\%    && 10\% & 30\%  \\
\end{scotch}
\vspace*{2ex}
\begin{scotch}{lcc{c}@{\hspace*{5pt}}cc}
\pPb & \multicolumn{2}{ c }{$n = 2$} &&  \multicolumn{2}{ c }{$n = 3$} \\
\cline{2-3}\cline{5-6}
\multicolumn{1}{l}{$N^{\text{offline}}_{\text{trk}}$}  & $\alpha=1$  & $\alpha=2$ &&  $\alpha=1$  & $\alpha=2$ \\
{[220, 260)}  & 1\% & 1.5\%  &&  1\%  & 20\%   \\
{[185, 220)}  & 1\% & 2.0\%  &&  1\% & 20\%   \\
{[150, 185)}  & 1\% & 2.0\%  &&  1\% & 20\%   \\
{[120, 150)}  & 1\% & 2.0\%  &&  1\% & 20\%   \\
\end{scotch}
\end{table}

\begin{figure*}
\centering
\includegraphics[width=0.95\textwidth]{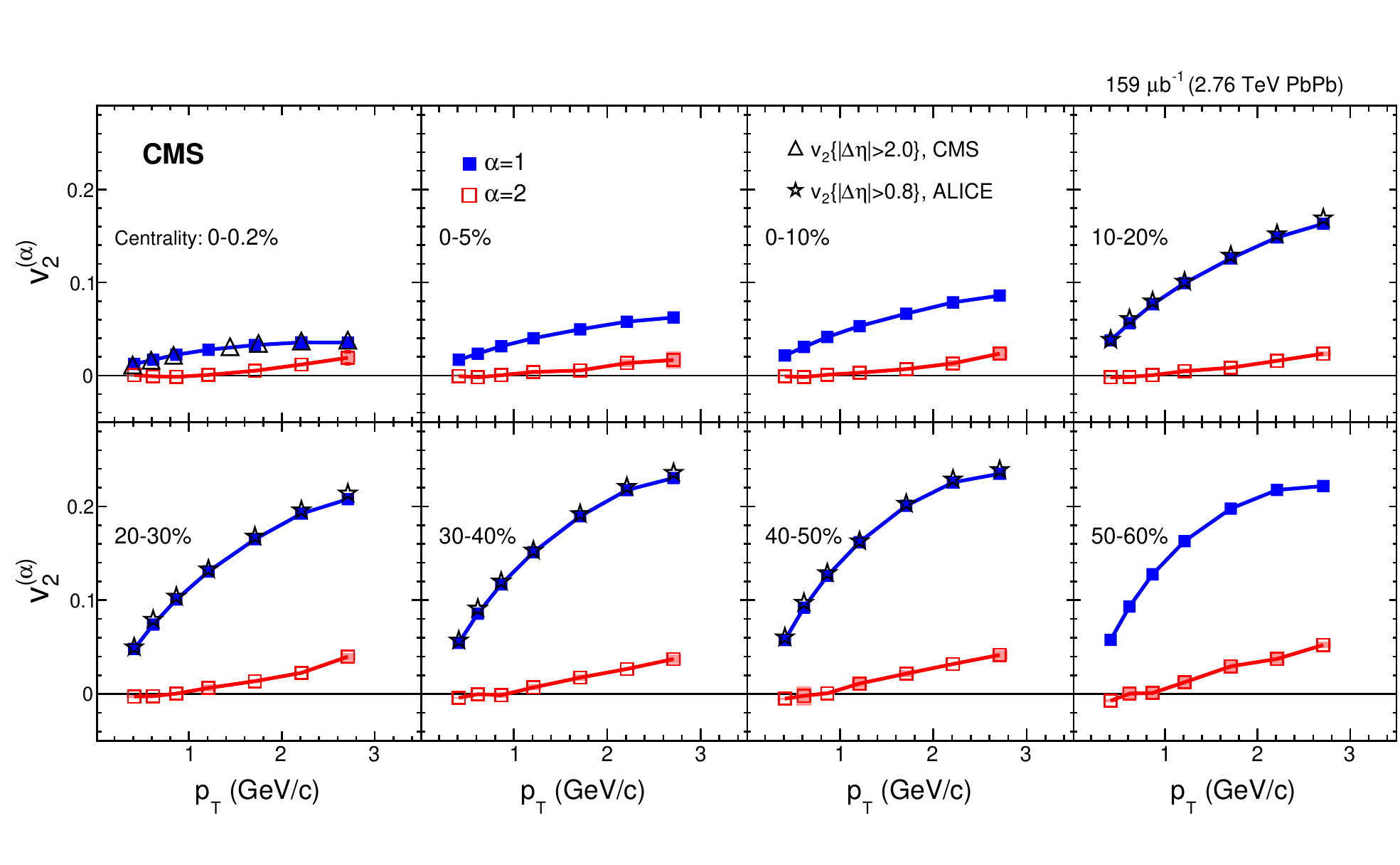}
\caption{ \label{fig:v2_PbPb}
Leading ($\alpha=1$) and subleading ($\alpha=2$) modes
for $n=2$ as a function of \pt, measured in a wide centrality range of \PbPb
collisions at $\rootsNN = 2.76$\TeV. The results for the leading mode
($\alpha=1$) are compared to the standard elliptic flow magnitude measured
by ALICE and CMS using the two-particle correlation method taken
from Refs. \cite{Aamodt:2011by, HIN_12_011}, respectively. The error bars correspond to
statistical uncertainties and boxes to systematic ones.
}
\end{figure*}
\begin{figure*}
\centering
\includegraphics[width=0.95\textwidth]{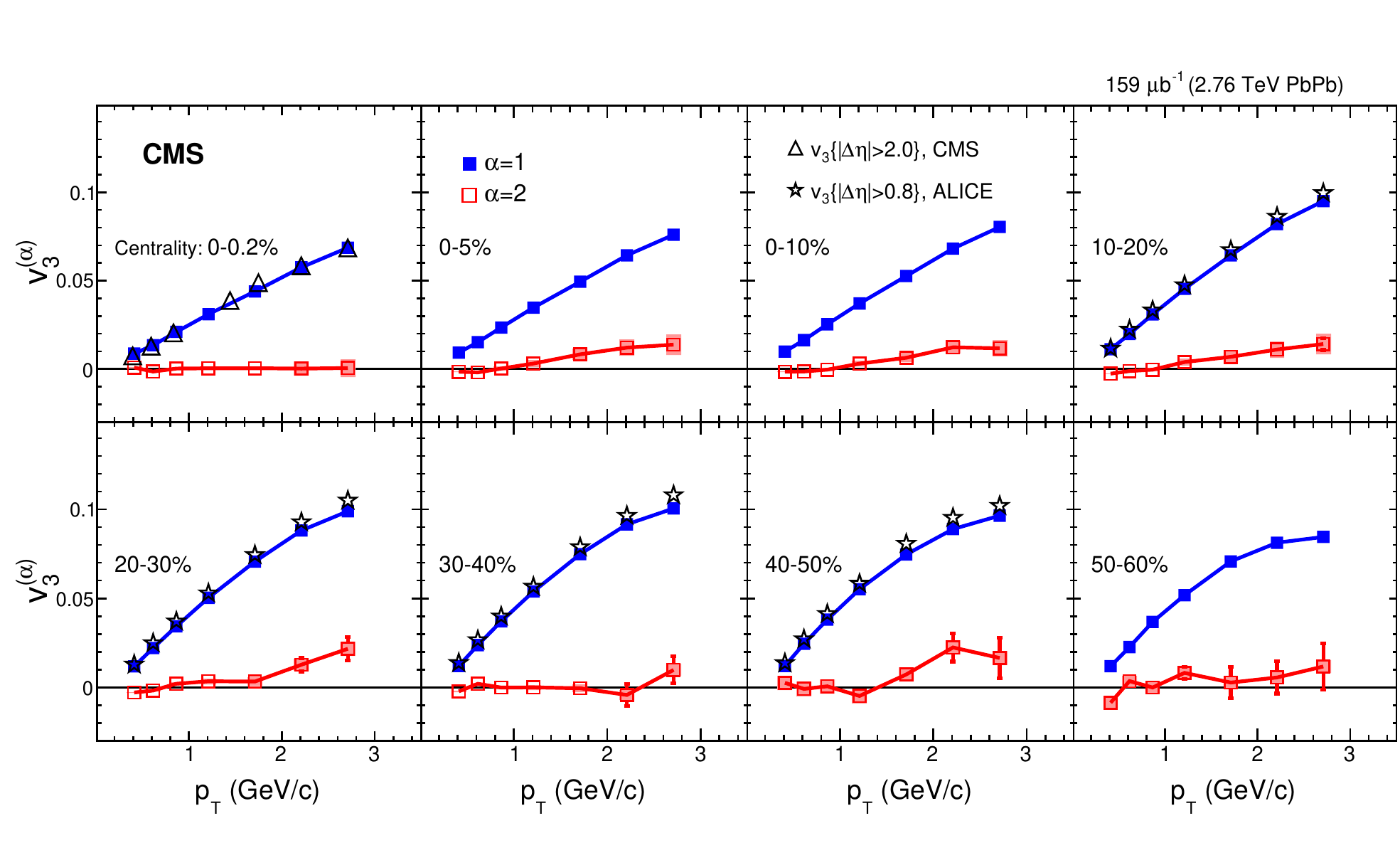}
\caption{ \label{fig:v3_PbPb}
Leading ($\alpha=1$) and subleading ($\alpha=2$) modes for $n=3$ as a function of
\pt, measured in a wide centrality range of \PbPb collisions at
$\rootsNN = 2.76$\TeV. The results for the leading mode ($\alpha=1$) are
compared to the standard triangular flow magnitude measured by
ALICE and CMS using the two-particle correlation method taken
from Refs. \cite{Aamodt:2011by, HIN_12_011}, respectively. The error bars correspond to
statistical uncertainties and boxes to systematic ones.
}
\end{figure*}
\begin{figure}
\centering
\includegraphics[width=\cmsFigWidth]{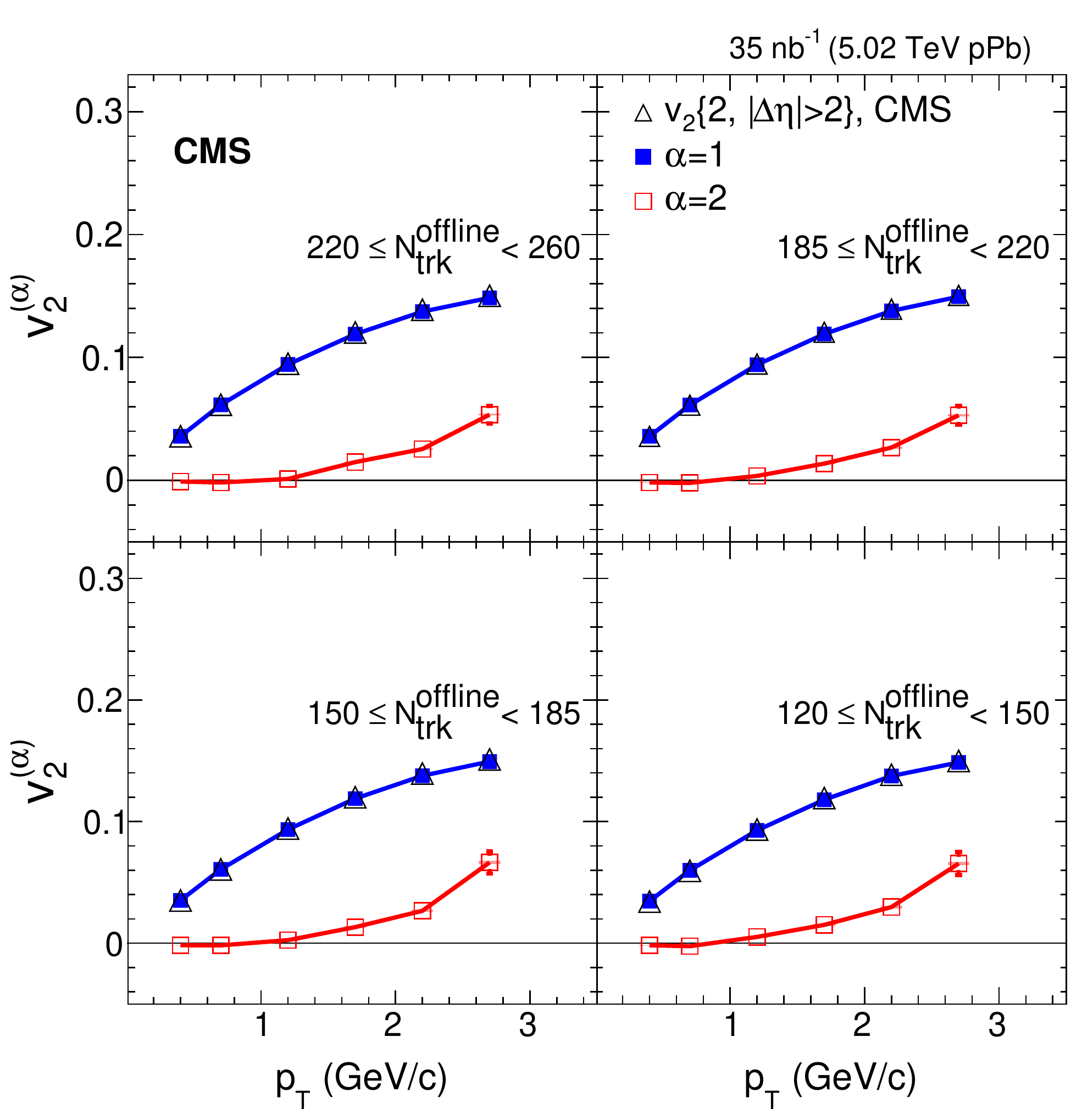}
\caption{ \label{fig:v2_pPb}
Leading ($\alpha=1$) and subleading ($\alpha=2$) modes for $n=2$ as a function of
\pt, measured in high-multiplicity \pPb collisions at
$\rootsNN = 5.02$\TeV, for four classes of reconstructed track multiplicity $N^\text{offline}_{\text{trk}}$. The results for the leading mode ($\alpha=1$) are
compared to the standard elliptic flow magnitude taken
from Ref. \cite{Chatrchyan:2013nka}.  The error bars correspond to
statistical uncertainties and boxes to systematic ones.
}
\end{figure}
\begin{figure} 
\centering
\includegraphics[width=\cmsFigWidth]{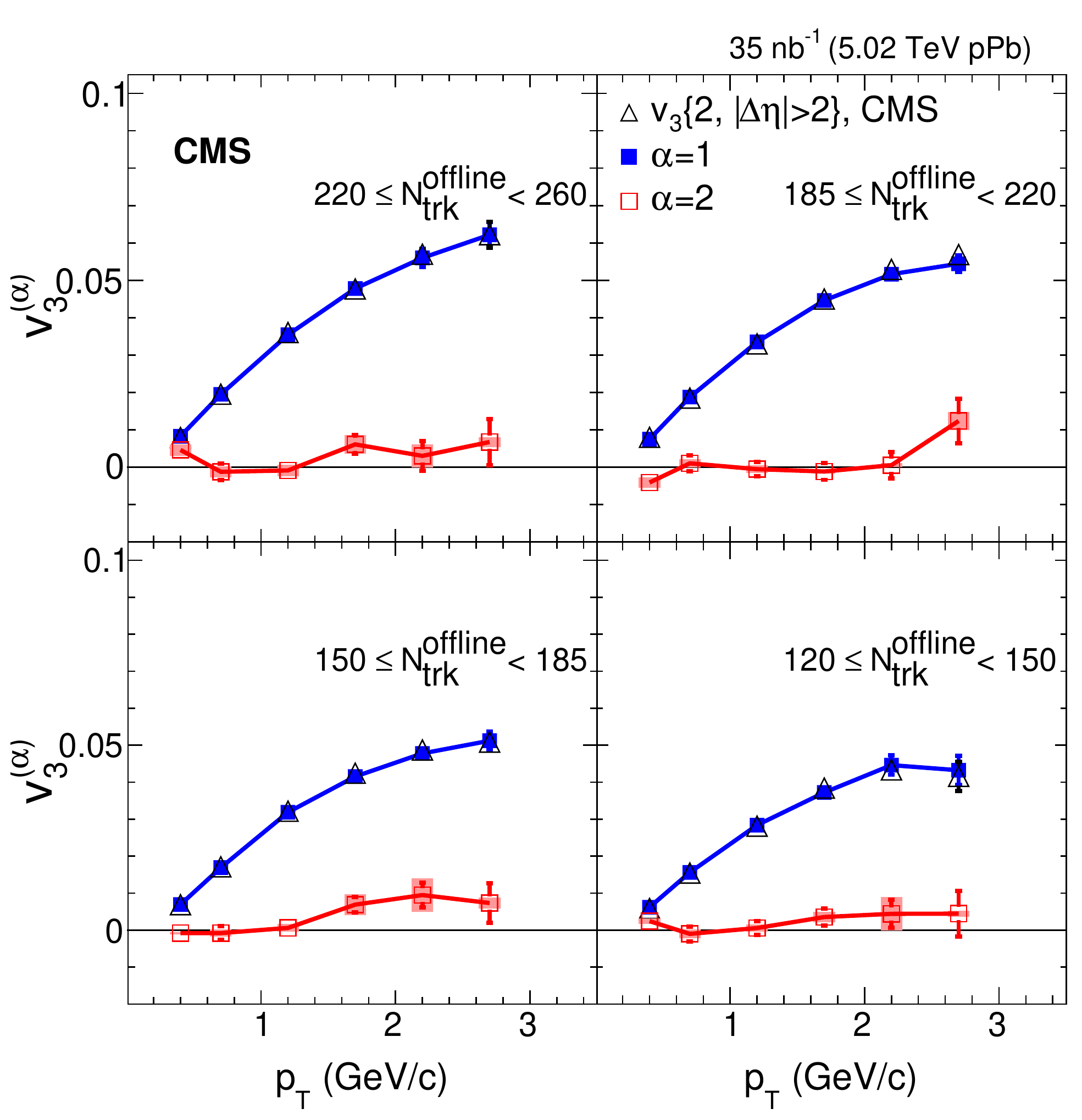}
\caption{ \label{fig:v3_pPb}
Leading ($\alpha=1$) and subleading ($\alpha=2$) modes for $n=3$ as a function of
\pt, measured in high-multiplicity \pPb collisions at
$\rootsNN = 5.02$\TeV, for four classes of reconstructed track multiplicity $N^\text{offline}_{\text{trk}}$. The results for the leading mode ($\alpha=1$) are
compared to the standard triangular flow magnitude taken
from Ref. \cite{Chatrchyan:2013nka}.  The error bars correspond to
statistical uncertainties and boxes to systematic ones.
}
\end{figure}
\begin{figure}
\centering
\includegraphics[width=1.0\linewidth]{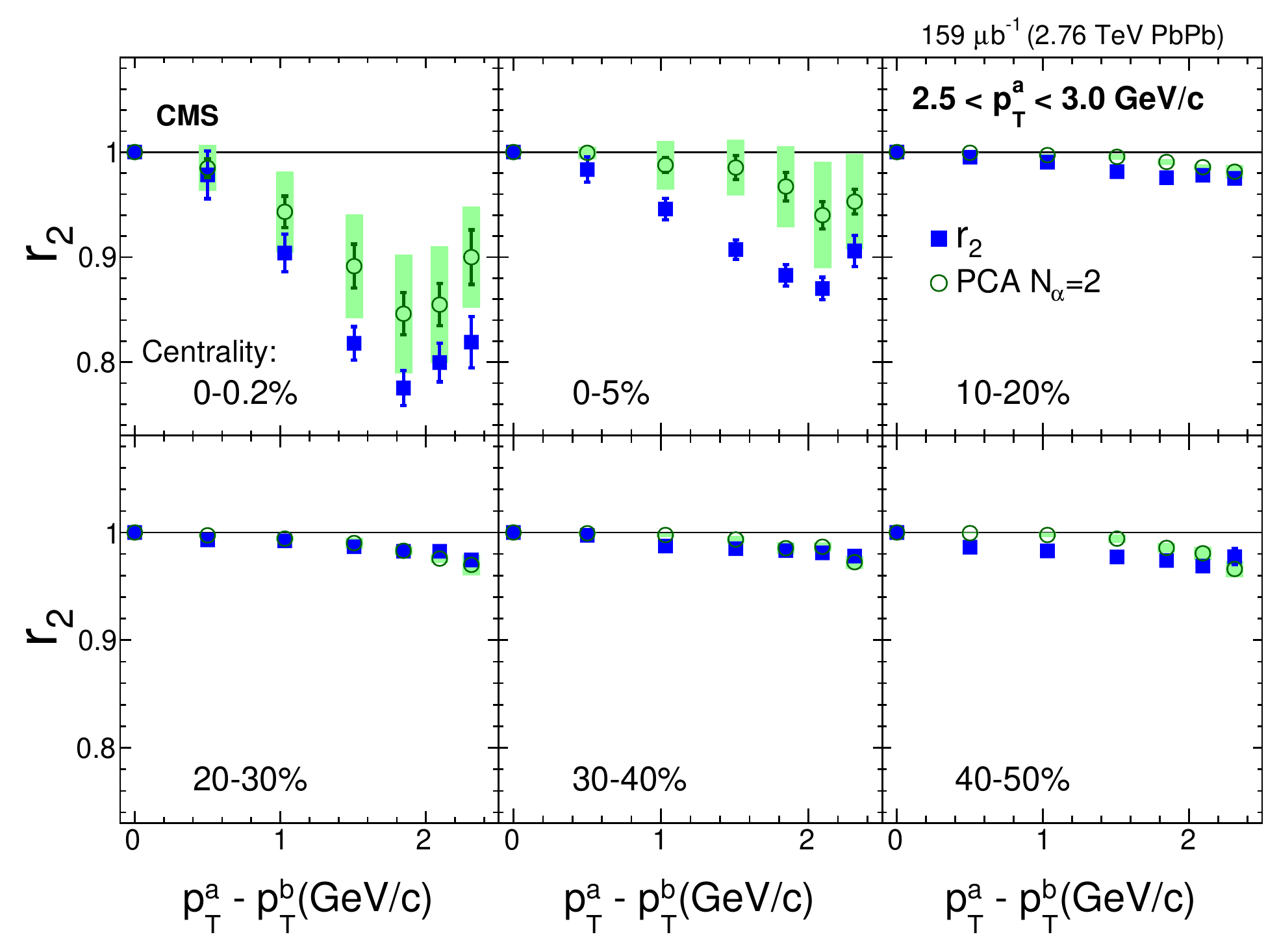}
\caption{ \label{fig:full_r2}
Comparison of the Pearson correlation coefficient $r_2$ reconstructed with harmonic decomposition using the leading and subleading modes and $r_2$ values from Ref.~\cite{Khachatryan:2015oea},
as a function of $\pt^a-\pt^b$ in bin of $\pt^a$ for six centrality classes
in \PbPb collisions at $\rootsNN = 2.76\TeV$. The error bars correspond to
statistical uncertainties and boxes to systematic ones.
}
\end{figure}
\begin{figure}
\centering
\includegraphics[width=1.0\linewidth]{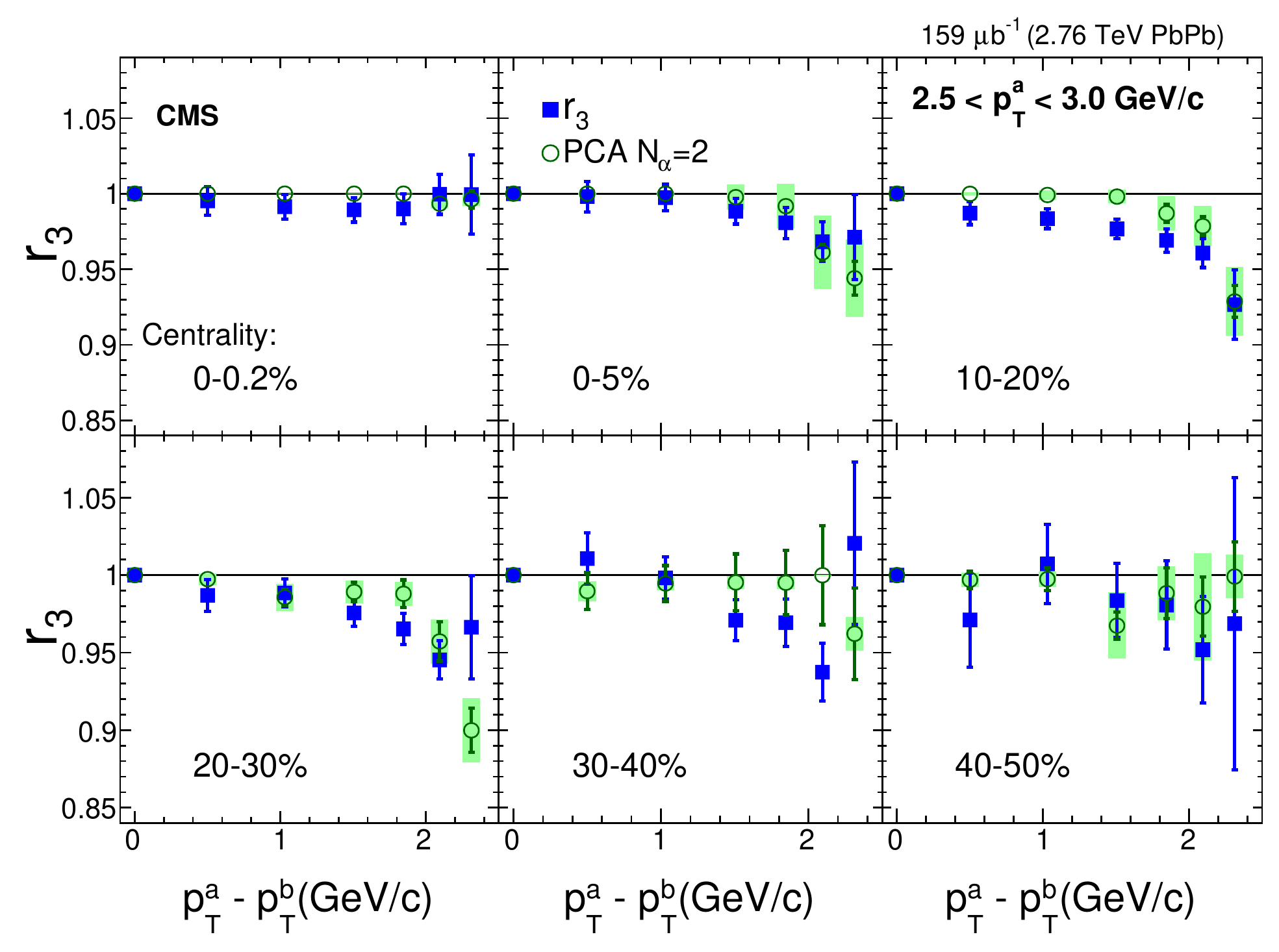}
\caption{ \label{fig:full_r3}
Comparison of the Pearson correlation coefficient $r_3$ reconstructed with harmonic decomposition using the leading and subleading modes and $r_3$ values from Ref.~\cite{Khachatryan:2015oea},
as a function of $\pt^a-\pt^b$ in bin of $\pt^a$ for six centrality classes
in \PbPb collisions at $\rootsNN = 2.76\TeV$. The error bars correspond to
statistical uncertainties and boxes to systematic ones.
}
\end{figure}
\begin{figure}
\centering
\includegraphics[width=1.0\linewidth]{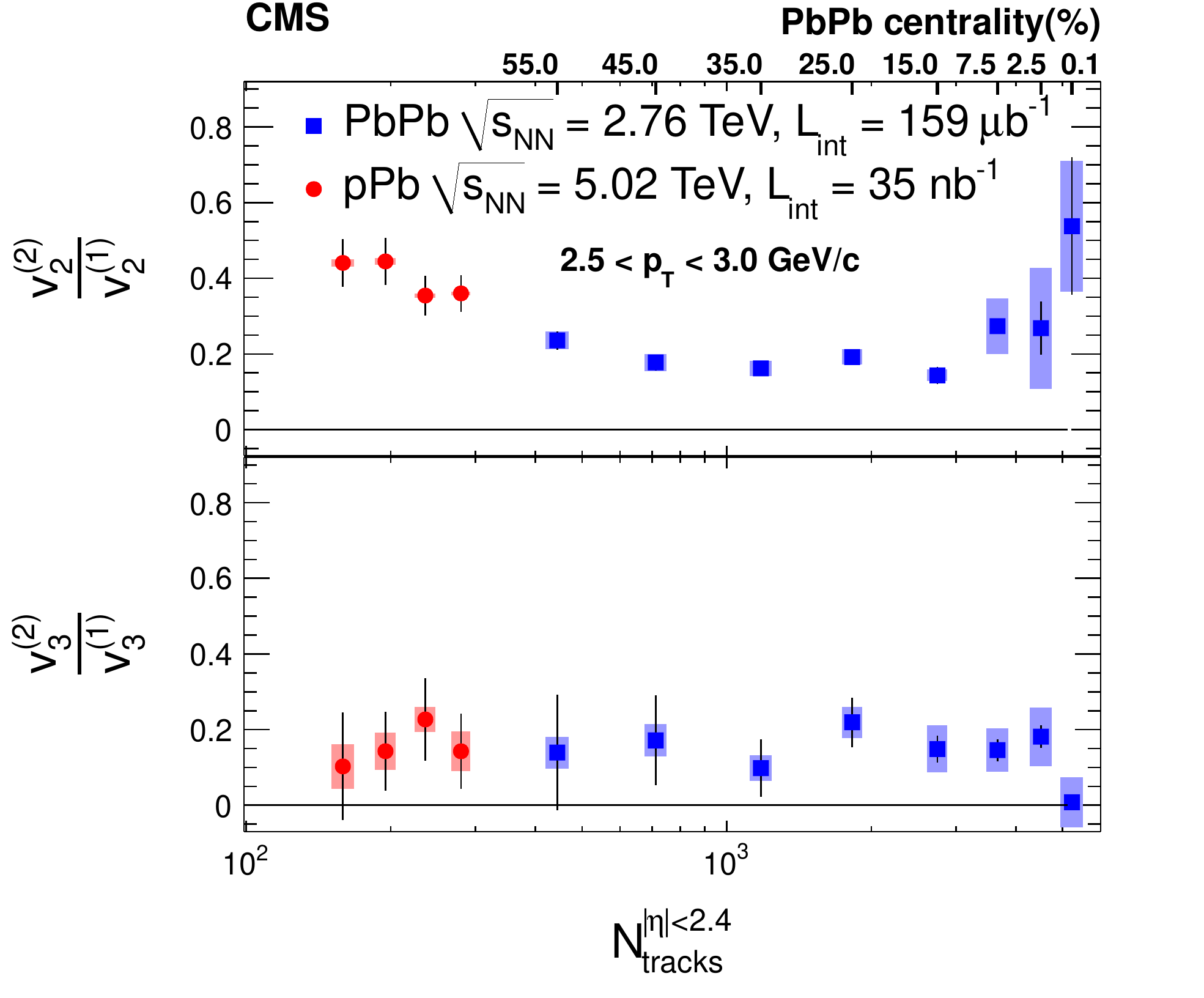}
\caption{ \label{fig:integrated_23_08}
The ratio between values of the subleading and leading modes, taken for the highest \pt bin, as a function of centrality and of
charged-particle multiplicity at midrapidity (double axis).
The PCA flow results for \PbPb collisions at $\rootsNN = 2.76$\TeV (filled blue squares) and for \pPb
collisions at $\rootsNN = 5.02$\TeV (filled red circles).
The error bars correspond to statistical uncertainties and boxes to systematic ones.
}
\end{figure}
\begin{figure*}
\centering
\includegraphics[width=0.95\textwidth]{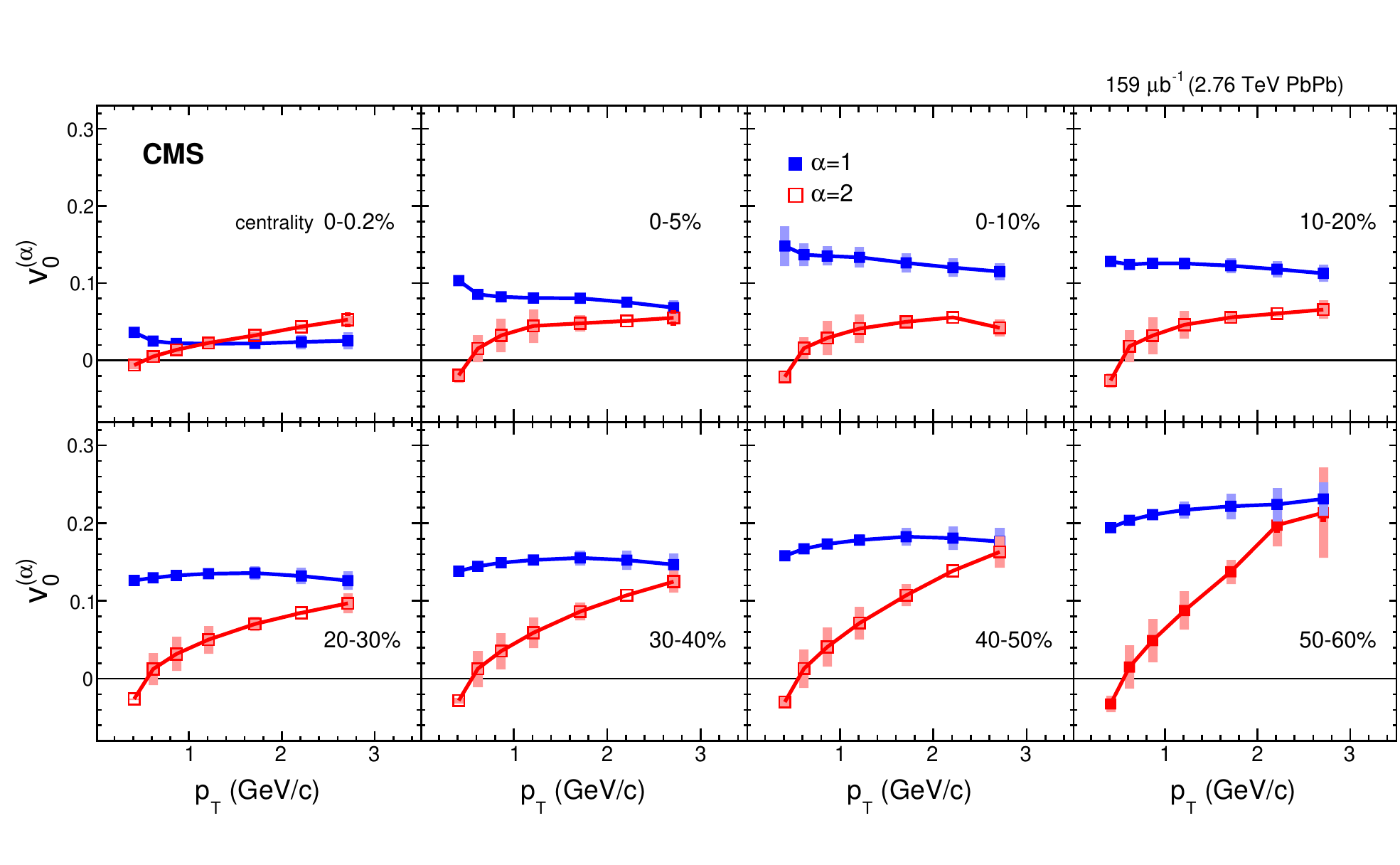}
\caption{ \label{fig:v0_PbPb}
Leading and subleading modes for $n=0$, \ie fluctuations in the total multiplicity, spanning eight centralities in \PbPb collisions at $\rootsNN = 2.76$\TeV.
The error bars correspond to
statistical uncertainties and boxes to systematic ones. The systematic uncertainties are strongly correlated bin-to-bin.
}
\end{figure*}

\section{Results}
\label{sec:results}

Figure \ref{fig:v2_PbPb} shows leading and subleading modes for the elliptic case ($n=2$)
for eight centrality regions in \PbPb collisions at $\rootsNN = 2.76$\TeV as a function of \pt.
These centrality regions range from ultracentral 0--0.2\% to peripheral 50--60\%.
The data are binned into seven \pt bins covering the region $0.3 <\pt< 3.0$\GeVc.
The number of differential \pt bins for constructing the covariance matrix is $N_\alpha=7$. In all the figures the points
are placed at the mean \pt value within a given bin.
For comparison, $v^{(1)}_2$ is plotted
together with $v_2\{2\}$  from CMS for ultracentral collisions \cite{HIN_12_011} and from ALICE for midcentral collisions \cite{Aamodt:2011by}.
The leading mode, $v^{(1)}_2$, is dominant and is
essentially equal to the single-particle anisotropy $v_2\{2\}$ extracted from two-particle correlations.
The subleading mode, $v^{(2)}_2$, is nonzero for all centrality classes and it tends to rise with \pt.
It has a small magnitude of about 0.02 for the highest \pt bin and more central collisions and then gradually increases up to 0.05 towards peripheral collisions.

Figure \ref{fig:v3_PbPb} shows leading and subleading modes
for the triangular case ($n=3$), using the same eight centrality classes in
PbPb collisions at $\rootsNN = 2.76$\TeV. Similar to the $n=2$ case, $v^{(1)}_3$ is plotted
together with $v_3\{2\}$ from CMS for ultracentral collisions \cite{HIN_12_011} and from ALICE for midcentral collisions \cite{Aamodt:2011by}.
A very good agreement is found between $v^{(1)}_3$ and the standard $v_3\{2\}$. The subleading mode,
$v^{(2)}_3$, is practically zero for ultracentral collisions but shows positive values for a range of centralities at high \pt.
From a hydrodynamical point of view the
existence of the subleading mode for $n=3$ is the response to the first radial excitation of triangularity \cite{PCA_PUB2}.

Figure \ref{fig:v2_pPb} shows leading
and subleading modes in the case of the elliptic
harmonic ($n=2$) in \pPb collisions at
$\rootsNN = 5.02$\TeV as a function of \pt for four different classes of multiplicity. The data are binned into six \pt bins covering the region $0.3<\pt<3.0$\GeVc.
The number of differential \pt bins for constructing the covariance matrix is $N_\alpha=6$. As seen in \PbPb collisions, the leading mode
is equal to standard $v_2\{2\}$ CMS results from Ref.~\cite{Chatrchyan:2013nka}.
Looking at the subleading mode ($\alpha=2$) values close to zero are observed at low \pt with a moderate increase in magnitude towards high
\pt. For \pt values close to 3.0\GeVc the subleading mode $\upsilon_{2}^{(2)}$ has a significant nonzero magnitude. This is the same \pt region where the
biggest factorization breaking has been seen in high-multiplicity \pPb collisions \cite{Khachatryan:2015oea}.
For both the leading and subleading elliptic modes, the data shows little multiplicity dependence for \pPb collisions.

Figure \ref{fig:v3_pPb} shows leading and
subleading modes for the triangular case ($n = 3$) for the same
multiplicity intervals from high-multiplicity \pPb collisions. As for the \PbPb case,
the differential values of the standard single-particle anisotropy $v_3\{2\}$ from Ref. \cite{Chatrchyan:2013nka} and
$v^{(1)}_3$ are equal.
The bottom panel of Fig. \ref{fig:v3_pPb} shows that $v^{(2)}_3$ is close to zero for all values of \pt. Quantitatively similar behavior was seen for flow factorization breaking in Ref. \cite{Khachatryan:2015oea}.
Similarly to the elliptic case, the leading and subleading triangular modes are rather independent of multiplicity for \pPb collisions.

The Pearson correlation coefficient defined in Eq.~(\ref{rn_def}) measures the magnitude of factorization breaking. This coefficient depends upon the two-particle harmonics $V_{n\Delta}$ that in turn are built up from the complete set of modes as
shown in Eq.~(\ref{sum_alpha}). These harmonics are approximated by the sum of just the leading and subleading modes.
 The comparison between the values of the PCA $r_2$ and of the $r_2$ from Ref. \cite{Khachatryan:2015oea} is shown in Fig. \ref{fig:full_r2}.
Using only the leading and subleading modes it is possible to reconstruct the shape of the $r_2$.
However $r_2$ is closer to unity for the PCA results than for the previous measurements. This is expected because the $V_{n\Delta}$
values are constructed from only two of the modes.
Figure \ref{fig:full_r3} shows the $n = 3$ case, again using the comparison with $r_3$ from the previous two-particle correlation analysis \cite{Khachatryan:2015oea}.
Although the errors are large it is clear that the principle-component analysis tracks the previously measured divergence of $r_3$ from unity at high \pt.

The Pearson coefficient calculated from Eq.~(\ref{rn_def})  can be expanded as a power series of ratios of modes.
Figure \ref{fig:integrated_23_08} shows the ratio of the leading and subleading modes for both \pPb and \PbPb collisions as a function of centrality
(track multiplicity). The ratios are calculated for the highest \pt bin used in the analysis. The top panel shows the elliptic case while the bottom panel shows the triangular case.
For the elliptic case the ratio is clearly above zero, with \pPb high-multiplicity values being above the peripheral \PbPb ones. For the triangular case half of the
individual points are consistent with zero within the uncertainties. However, the ensemble of all the points
suggest that the ratio is above zero.

Finally, Fig. \ref{fig:v0_PbPb} shows leading and subleading modes for the multiplicity case ($n=0$) for \PbPb collisions as a
function of \pt for eight regions of  centrality. For all centralities the leading mode depends only weakly on \pt, while the subleading mode increases rapidly with \pt except for very central collisions.
The observed increase of the subleading mode with \pt for all centralities is a response to radial-flow fluctuations \cite{PCA_PUB1, pt_spectra}.
From a hydrodynamical point of view, the number of particles at high-\pt decreases exponentially as $\exp{[\pt (u-u_0 )/T ]}$.
Here, $T$ is the temperature, $u$ is the maximum fluid velocity, and $u_0=\sqrt{1+u^2}$. A small variation in $u$ produces a relative yield that increases linearly with \pt. Such behaviour
is observed in the data for more peripheral collisions.
At a given \pt the subleading mode increases strongly from central to peripheral collisions. Since peripheral collisions correspond to smaller interaction volumes, it is
expected that \pt fluctuations are more important for peripheral than for central events.

\section{Summary}
\label{sec:conclusion}

For the first time the leading and subleading modes of elliptic and triangular flow have been
measured for 5.02\TeV \pPb and 2.76\TeV \PbPb collisions. For \PbPb collisions the leading and
subleading modes of multiplicity fluctuations have also been measured. Since the principal
component analysis uses all the information encoded in the covariance matrix, it provides
increased sensitivity to fluctuations.
For a very wide range of \pt and centrality, the leading
modes of the elliptic and triangular flow are found to be essentially equal to the anisotropy coefficients  measured using the standard two-particle correlation method. For both the elliptic and triangular cases the subleading modes are non-zero and  increase with \pt. This behavior reflects a
breakdown of flow factorization at high \pt in both the \pPb and \PbPb systems.
For charged-particle multiplicity both the leading and subleading modes
 increase steadily from central to peripheral \PbPb events. The leading mode
depends only weakly upon \pt  while the  subleading
 mode increases strongly with \pt. This centrality and \pt dependence is  suggestive of the presence of fluctuations in the radial flow.

In summary the subleading modes of the principal-component analysis  capture new information from
the spectra of flow and multiplicity fluctuations and provide an efficient method to quantify the breakdown of factorization in two-particle correlations.

\begin{acknowledgments}
We congratulate our colleagues in the CERN accelerator departments for the excellent performance of the LHC and thank the technical and administrative staffs at CERN and at other CMS institutes for their contributions to the success of the CMS effort. In addition, we gratefully acknowledge the computing centers and personnel of the Worldwide LHC Computing Grid for delivering so effectively the computing infrastructure essential to our analyses. Finally, we acknowledge the enduring support for the construction and operation of the LHC and the CMS detector provided by the following funding agencies: BMWFW and FWF (Austria); FNRS and FWO (Belgium); CNPq, CAPES, FAPERJ, and FAPESP (Brazil); MES (Bulgaria); CERN; CAS, MoST, and NSFC (China); COLCIENCIAS (Colombia); MSES and CSF (Croatia); RPF (Cyprus); SENESCYT (Ecuador); MoER, ERC IUT, and ERDF (Estonia); Academy of Finland, MEC, and HIP (Finland); CEA and CNRS/IN2P3 (France); BMBF, DFG, and HGF (Germany); GSRT (Greece); OTKA and NIH (Hungary); DAE and DST (India); IPM (Iran); SFI (Ireland); INFN (Italy); MSIP and NRF (Republic of Korea); LAS (Lithuania); MOE and UM (Malaysia); BUAP, CINVESTAV, CONACYT, LNS, SEP, and UASLP-FAI (Mexico); MBIE (New Zealand); PAEC (Pakistan); MSHE and NSC (Poland); FCT (Portugal); JINR (Dubna); MON, RosAtom, RAS, RFBR and RAEP (Russia); MESTD (Serbia); SEIDI, CPAN, PCTI and FEDER (Spain); Swiss Funding Agencies (Switzerland); MST (Taipei); ThEPCenter, IPST, STAR, and NSTDA (Thailand); TUBITAK and TAEK (Turkey); NASU and SFFR (Ukraine); STFC (United Kingdom); DOE and NSF (USA).

\hyphenation{Rachada-pisek} Individuals have received support from the Marie-Curie program and the European Research Council and Horizon 2020 Grant, contract No. 675440 (European Union); the Leventis Foundation; the A. P. Sloan Foundation; the Alexander von Humboldt Foundation; the Belgian Federal Science Policy Office; the Fonds pour la Formation \`a la Recherche dans l'Industrie et dans l'Agriculture (FRIA-Belgium); the Agentschap voor Innovatie door Wetenschap en Technologie (IWT-Belgium); the Ministry of Education, Youth and Sports (MEYS) of the Czech Republic; the Council of Science and Industrial Research, India; the HOMING PLUS program of the Foundation for Polish Science, cofinanced from European Union, Regional Development Fund, the Mobility Plus program of the Ministry of Science and Higher Education, the National Science Center (Poland), contracts Harmonia 2014/14/M/ST2/00428, Opus 2014/13/B/ST2/02543, 2014/15/B/ST2/03998, and 2015/19/B/ST2/02861, Sonata-bis 2012/07/E/ST2/01406; the National Priorities Research Program by Qatar National Research Fund; the Programa Clar\'in-COFUND del Principado de Asturias; the Thalis and Aristeia programs cofinanced by EU-ESF and the Greek NSRF; the Rachadapisek Sompot Fund for Postdoctoral Fellowship, Chulalongkorn University and the Chulalongkorn Academic into Its 2nd Century Project Advancement Project (Thailand); and the Welch Foundation, contract C-1845.
\end{acknowledgments}
\clearpage
\bibliography{auto_generated}
\cleardoublepage \appendix\section{The CMS Collaboration \label{app:collab}}\begin{sloppypar}\hyphenpenalty=5000\widowpenalty=500\clubpenalty=5000\textbf{Yerevan Physics Institute,  Yerevan,  Armenia}\\*[0pt]
A.M.~Sirunyan, A.~Tumasyan
\vskip\cmsinstskip
\textbf{Institut f\"{u}r Hochenergiephysik,  Wien,  Austria}\\*[0pt]
W.~Adam, F.~Ambrogi, E.~Asilar, T.~Bergauer, J.~Brandstetter, E.~Brondolin, M.~Dragicevic, J.~Er\"{o}, M.~Flechl, M.~Friedl, R.~Fr\"{u}hwirth\cmsAuthorMark{1}, V.M.~Ghete, J.~Grossmann, J.~Hrubec, M.~Jeitler\cmsAuthorMark{1}, A.~K\"{o}nig, N.~Krammer, I.~Kr\"{a}tschmer, D.~Liko, T.~Madlener, I.~Mikulec, E.~Pree, D.~Rabady, N.~Rad, H.~Rohringer, J.~Schieck\cmsAuthorMark{1}, R.~Sch\"{o}fbeck, M.~Spanring, D.~Spitzbart, J.~Strauss, W.~Waltenberger, J.~Wittmann, C.-E.~Wulz\cmsAuthorMark{1}, M.~Zarucki
% ...
\vskip\cmsinstskip
% ...
1:~~Also at Vienna University of Technology, Vienna, Austria\\
% ...

\end{sloppypar}
\end{document}